\documentclass{PoS}

\usepackage{amsmath,amstext,amsfonts,amsbsy,amssymb,amscd,bbm,epsfig,lscape}

\newcommand{\ba}{\begin{array}}
\newcommand{\ea}{\end{array}}

\newcommand{\refig}[1]{Fig.~\ref{#1}}

\newcommand{\dif}{{\rm d}}
\newcommand{\Tr}{{\rm Tr}}

\newcommand{\Dslash}{\relax{\kern+.25em / \kern-.70em D}}

\newcommand{\fm}{{\rm fm}}
\newcommand{\MeV}{{\rm MeV}}

\newcommand{\Real}{\relax{\mathsf{\Gamma\kern-.35em R}}}
\newcommand{\Int}{\relax{\mathsf{Z\kern-.40em Z}}}

\newcommand{\half}{{\scriptstyle{{1\over 2}}}}

\newcommand{\twth}{{\scriptstyle{{1\over 12}}}}

\newcommand{\gbar}{\kern1pt\overline{\kern-1pt g\kern-0pt}\kern1pt}
\newcommand{\mbar}{\kern2pt\overline{\kern-1pt m\kern-1pt}\kern1pt}
\newcommand{\obar}[1]{\kern3pt\overline{\kern-2pt #1\kern-0pt}\kern1pt}

\newcommand{\Oa}{\mbox{O}(a)}
\newcommand{\Oasq}{\mbox{O}(a^2)}

\newcommand{\abar}{\kern1pt\overline{\kern-1pt a\kern-0.5pt}\kern1pt}

\newcommand{\cC}{{\cal C}}

\newcommand{\cJ}{{\cal J}}

\newcommand{\cL}{{\cal L}}

\newcommand{\vn}{\mathbf{n}}

\title{Mixed action computations on fine dynamical lattices\thanks{IFIC/09-60,FTUV-09-1119,FTUAM-09-32,IFT-UAM/CSIC-09-57,CERN-PH-TH-2009-222.}}

\ShortTitle{Mixed action computations on fine dynamical lattices}

\author{F.~Bernardoni, P.~Hern\'andez\\
        Instituto de F\'{\i}sica Corpuscular, CSIC-Universitat de Val\`encia\\
		Apartado de Correos 22085, E-46071 Valencia, Spain\\
        E-mail: \email{fabio.bernardoni@ific.uv.es},~\email{pilar.hernandez@ific.uv.es}}

\author{N.~Garron\thanks{Address after Nov 1: SUPA, School of Physics and Astronomy, Univ. of Edinburgh, Edinburgh EH9 3JZ, UK.}\\
        Instituto de F\'{\i}sica Te\'orica UAM/CSIC\\
        Universidad Aut\'onoma de Madrid, Cantoblanco E-28049 Madrid, Spain\\
        E-mail: \email{nicolas.garron@uam.es}}

\author{S.~Necco\\
        CERN, Physics Department, TH Division\\
        CH-1211 Geneva 23, Switzerland\\
        E-mail: \email{necco@mail.cern.ch}}

\author{\speaker{C.~Pena}\\
        Dpto. de F\'{\i}sica Te\'orica and Instituto de F\'{\i}sica Te\'orica UAM/CSIC\\
        Universidad Aut\'onoma de Madrid, Cantoblanco E-28049 Madrid, Spain\\
        E-mail: \email{carlos.pena@uam.es}}

\abstract{We report on our first experiences in simulating Neuberger valence fermions on CLS $N_f=2$ configurations with light sea quark masses and small lattice spacings. Valence quark masses are considered that allow to explore the matching to (partially quenched) chiral perturbation theory both in the $\epsilon$- and $p$-regimes. The setup is discussed, and first results are presented for spectral observables.}

\FullConference{The XXVII International Symposium on Lattice Field Theory\\
		 July 26-31, 2009\\
		 Peking University, Beijing, China}

\begin{document}

\vspace{-2mm}\section{Introduction}\vspace{-2mm}

Thanks to the theoretical and algorithmic improvements of recent years,
and to the ever increasing computational power available,
state-of-the-art Lattice QCD simulations now routinely take place
at dynamical pion masses in the 200--300~MeV ballpark~\cite{dynrev}. In this mass region 
the effective description of the dynamics of pseudo-Goldstone bosons
at low energies by means of chiral perturbation theory (ChiPT) is expected to work
well at a quantitative level. This gives rise to a fertile interaction: by matching
Lattice QCD and ChiPT results it is possible, on the one hand, to test the effective
description vs. the fundamental theory; and, on the other hand,
low-energy constants (LECs) can be determined from first principles, thus providing
a sounder foundation to phenomenological applications of ChiPT.

One particularly interesting aspect of the matching between QCD and ChiPT is
the role of finite volume effects~\cite{epsilon}
(we will always assume that the theory lives in
an Euclidean four-volume $V=L^3 \times T$). While for large enough values of
$L$ (one typical estimate is $m_\pi L \gtrsim 4$) the latter are expected
to be strongly suppressed, and give rise only to small corrections to the infinite
volume expansion in powers of pion momenta, the situation changes completely when the 
Compton wavelength of pions approaches $L$, i.e. $m_\pi L \sim 1$. In this
regime slow pion modes, strongly affected by the finite volume, dominate the path
integral in the effective theory, and the expansion in powers of $p^2/\Lambda_\chi^2$
breaks down. Indeed, the resulting finite volume chiral regimes involve a
rearrangement of the chiral expansion, in which mass effects are suppressed relative to 
volume effects; as a consequence, less LECs appear at any given order in 
the expansion relative to the infinite volume case. This in turn leads to a very
different setup for the determination of LECs, which offers both the potential
to obtain cleaner computations of some of the latter (those whose effects are unsuppressed in the quark mass),
and a cross-check of the systematic uncertainties of ``infinite''-volume studies.

Another key property of finite volume chiral regimes is that the low-lying spectrum of the Dirac operator can be described by an appropriate random
matrix theory (RMT)~\cite{rmt}. Direct quantitative tests of such description
have already been obtained both in quenched~\cite{rmtnum1} and $N_f=2$ QCD~\cite{rmtnum2}. Since RMT predictions
are sensitive to the value of the chiral condensate, they provide yet another way
of studying chiral symmetry breaking, using simple spectral observables.

Obviously enough, an adequate treatment of chiral symmetry on the lattice 
is especially relevant in this context. While simulations of $N_f=2(+1)$ QCD with
full chiral symmetry have proven feasible, they are still limited to relatively small
values of the inverse lattice spacing and/or physical volume~\cite{dynov}.
A way to overcome this is
to use a mixed action approach, in which chiral symmetry is exactly preserved at the
level of valence quarks only. Our aim is to develop such a framework by considering
Neuberger valence quarks on top of $N_f=2$ CLS ensembles, obtained from simulations
with non-perturbatively $\Oa$ improved Wilson sea quarks. A key ingredient of our
study will be the matching of QCD to ChiPT in a mixed regime, in which sea quark
masses are in the $p$-regime and valence quark masses can take values both in the
$p$- and the $\epsilon$-regime~\cite{mixedreg}.

Here we will report on
our first experiences with this approach, involving technical tests and finite
volume regime studies along the lines hinted at above.
We present results for spectral observables, which provide information on $\Sigma$ and $L_6$.
Obviously, mixed actions also have huge potential for
phenomenological applications in which the
exact preservation of chiral symmetry is greatly advantageous, e.g. to simplify
the renormalisation of composite operators entering hadronic matrix elements.
Along this line, first data for standard two- and three-point functions, as well as
for correlation
functions computed in the chiral limit via saturation with
topological zero modes~\cite{zm}, will be covered in upcoming publications.

\vspace{-2mm}\section{Probing the deep chiral regime with mixed actions}

\vspace{-2mm}\subsection{Mixed chiral regimes}\vspace{-1mm}

While the exploration of finite volume regimes ideally involves simulations with 
extremely light sea quarks, it is still possible to access them in a situation in
which sea pions have moderately large masses. The main idea is to formulate ChiPT
in a so-called mixed regime~\cite{mixedreg}, in which $N_h$ quarks have masses such that the
$p$-regime requirement $m_h\Sigma V\gg 1$ is satisfied, while $N_l$ quarks have masses
that fulfill the $\epsilon$-regime condition $m_l\Sigma V \lesssim 1$. The chiral
expansion proceeds by treating the Goldstone modes associated to the $N_h$ heavier
quarks essentially as decoupling particles.
In this way, NLO expressions for meson correlators in the light channel preserve
their typical $\epsilon$-regime features, with some extra terms (involving NLO LECs)
induced by loops involving heavier modes. Finally, it is possible to quench one of the
sectors, and extend the results for the full theory with the techniques of partially
quenched ChiPT.
Indeed, we will be interested in considering a quenched (valence) $\epsilon$-regime sector and a 
dynamical $p$-regime sector. When matching the chiral effective theory to QCD the valence 
sector of the latter will involve Neuberger quarks, while the sea sector will
contain non-perturbatively $\Oa$ improved Wilson $N_f=2$ CLS configurations.

Let us consider e.g. the NLO partially quenched
ChiPT prediction for the two-point function of (the zeroth component of)
a left-handed current $\cJ_0$ involving light quarks. We start from the LO chiral 
Lagrangian
\begin{gather}
\cL_{\rm ChiPT}=\frac{F^2}{4}\Tr\left[\partial_\mu U\partial_\mu U^\dagger\right]
-\frac{\Sigma}{2}\Tr\left[e^{i\theta} MU+U^\dagger M e^{-i\theta}\right]\,
\end{gather}
and the left handed chiral current defined as
$\cJ_\mu^a=(F^2/2)\Tr[T^aU\partial_\mu U^\dagger]$, with $T^a$ a flavour generator.
Its two-point function can be written as
\begin{gather}
\Tr[T^aT^b]\cC(x_0)=\int\dif^3x\left\langle\cJ_0^a(x)\cJ_0^b(0)\right\rangle\,.
\end{gather}
We consider observables defined at a fixed value of the 
topological charge $\nu$.
The NLO result for $\cC$ assuming the specific kinematics we are interested in
can be found in~\cite{mixedreg}
\begin{gather}
\label{eq:mixed_2p}
\cC(x_0) = \frac{F^2}{2T}\left\{
1-\frac{N_l}{F^2}\left[G(0,M_{hh}^2/2)-8L_4M_{hh}^2+\frac{T}{L^3}k_{00}^h\right]
+\frac{2T}{F^2L^3}\tilde\sigma_\nu(m_l,m_h)h_1(x_0/T)\right\}\,.
\end{gather}
In this expression $M_{ab}^2=(m_a+m_b)\Sigma/F^2$ is the mass of a $p$-regime pseudo
Nambu-Goldstone boson; the Green function $G$ is defined as
$G(x,M^2)=V^{-1}\sum_{n\in\mathbb{Z}^4}e^{ipx}(p^2+M^2)^{-1}$, with momenta
$p=2\pi(n_0/T,\vn/L)$;
$k_{00}^h$ is a constant that depends on the box geometry only;
$h_1(\tau)=\half[(\tau-\half)^2-\twth]$; and $\tilde\sigma_\nu$
is an explicitly known function of the dimensionless variables $m\Sigma V$, that
depends on the topology.
As announced, eq.~(\ref{eq:mixed_2p}) displays both the characteristic $\epsilon$-regime
parabolic dependence on $x_0$, and a characteristic $p$-regime NLO correction proportional
to $L_4$. Interestingly, Eq.~(\ref{eq:mixed_2p}) can be rewritten in the same form as in 
the $\epsilon$-regime quenched case, provided that the LEC $F$ is
replaced by the ``effective'' LEC
\begin{gather}
\tilde F^2 = F^2\left\{1-\frac{N_h}{F^2}\left[G_\infty(0,M_{hh}^2/2)-8L_4M_{hh}^2\right]\right\}\,,
\end{gather}
where $G_\infty$ is the infinite volume closed propagator, obtained by substituting the sum
over momenta in the Green function for an integral.\footnote{This implies
matching the mixed-regime expansion to a quenched effective theory, which in turn
involves a number of subtleties, mainly related to the treatment of mixed
heavy-light modes.
A detailed discussion can be found in~\cite{mixedreg}.}

The bottomline of this analysis is that the current two-point function in QCD,
computed in a mixed action framework with $\epsilon$-regime valence quarks and
$p$-regime sea quarks, is expected to exhibit a quenched $\epsilon$-regime form,
which can be fitted for $\tilde F$; by computing this effective LEC at a number
of sea pion masses $M_{hh}^2$ it is then possible to extract $F$ and $L_4$.
It has to be stressed that the LECs thus computed are the ones of the $N_f=2$
theory.

Similar results can be shown to hold for other LECs, rendering the strategy
general. For instance, the corresponding effective chiral condensate has the form
\begin{gather}
\label{eq:condensate}
\tilde\Sigma_r = \Sigma\left[
1-\frac{N_h}{F^2}\left[G_\infty(0,M_{hh}^2/2)-16L_6M_{hh}^2\right]+\frac{E_\infty}{F^2}
\right]\,,
\end{gather}
where $E_\infty$ is the well-known singlet contribution related to the renormalisation
of $\Sigma$ in the quenched theory. Eq.~(\ref{eq:condensate})
is particularly useful, as it can be used to extract information about LECs from
spectral observables, as will be discussed below.

\vspace{-2mm}\subsection{Random matrix theory}\vspace{-1mm}

It is well known that at the leading order of the chiral $\epsilon$-expansion
the partition function of ChiPT coincides, at any fixed value of the topological charge, 
with that of an appropriate chiral Random Matrix Theory describing the probability
distributions of the eigenvalues of the Dirac operator~\cite{rmt}.
RMT provides explicit predictions for the probability distributions $p_k(\zeta_k;\mu)$,
where $\lambda_k=\zeta_k/(\Sigma V)$ is the $k$-th eigenvalue of the massless Dirac 
operator and $\mu=m\Sigma V$, where $m$ is to be interpreted as a (small) sea quark
mass. $p_k$ depends on the number of dynamical
flavours and the topological charge $\nu$ through the combination $\xi=N_f+|\nu|$.

This establishes a direct
connection between the spectrum of the Dirac operator and the effective description
of QCD at low energies. As RMT provides an extremely detailed description of spectral
observables, such connection has an enormous potential as a tool to improve our
understanding of the QCD/ChiPT matching.
Of course, there is need of quantitative studies that check the extent of corrections
to RMT predictions, as those already performed in~\cite{rmtnum1,rmtnum2}. One of our purposes is to extend these analyses to larger
physical volumes and closer to the continuum limit. Also, we intend to explore the
potential of spectral observables to determine chiral LECs.
Of particular interest to us is the matching of spectral QCD results
to RMT in a mixed regime, in which eigenvalues of the Neuberger-Dirac operator are
computed on $N_f=2$ configurations with $p$-regime dynamical pion masses.

If sea pions were in the $\epsilon$-regime,
the results for $\langle\lambda_k\rangle_\nu$
(where $\langle\rangle_\nu$ stands for expectation values in a fixed topological
sector) are expected to match the RMT results for $\xi=2+|\nu|$.
If, on the other hand, the sea pion mass is large enough the theory will
approach quenched QCD, and
RMT should be worked out at $\xi=|\nu|$. One therefore
expects that $\langle\lambda_k\rangle_\nu$ displays a sea quark mass 
dependence that interpolates between both extremes. Remarkably, RMT does provide
a formula that interpolates smoothly between the $N_f=2$ and $N_f=0$ cases,
via the $\mu$ dependence of the probability distribution $p_k$.
It is however unclear how this $\mu$ dependence should be interpreted in the
transition region between the $\epsilon$- and $p$-regimes in sea quark masses, as in that case it is conceivable that spectral observables may receive sizeable corrections at
NLO in ChiPT, over which RMT has in principle no control. The results obtained in~\cite{rmtnum2} in this regard are inconclusive.

Indeed, a better grasp on the sea quark mass dependence can be obtained from a
matching to mixed-regime ChiPT: the LO partition function with $p$-regime sea pions
is that of a quenched theory in the $\epsilon$-regime, 
with a sea pion mass-dependent value of $\Sigma$ given by Eq.~(\ref{eq:condensate}).
This provides definite predictions for the sea mass dependence of
$\langle\lambda_k\rangle_\nu$. Consider e.g.
ratios of average eigenvalues of the form
$\langle\lambda_k\rangle_\nu(M_1)/ \langle\lambda_k\rangle_\nu(M_2)$, 
($M_{1,2}$ are two different sea pion masses). If we match
$\langle\lambda_k\rangle_\nu$ to quenched $\epsilon$-regime ChiPT
(i.e. quenched RMT) we expect the effective theory to
work with appropriate values $\tilde\Sigma_r(M_{1,2})$ of the effective chiral
condensate. Now, assuming no corrections to the RMT predicion other than this
mass dependence (which is consistent with our expansion scheme), we have
\begin{gather}
\label{eq:ev_condensate}
\frac{\langle\lambda_k\rangle_\nu(M_1)}{\langle\lambda_k\rangle_\nu(M_2)}
= \frac{\langle\zeta_k\rangle_{\nu,{\rm RMT}}}{\langle\zeta_k\rangle_{\nu,{\rm RMT}}}\,
\frac{\tilde\Sigma_r(M_2)}{\tilde\Sigma_r(M_1)}
=\frac{\tilde\Sigma_r(M_2)}{\tilde\Sigma_r(M_1)}\,.
\end{gather}
It follows that information on the mass dependence of $\tilde\Sigma_r$,
and hence on $L_6$, can be obtained from suitable eigenvalue ratios.

\vspace{-2mm}\section{Results on Dirac spectral observables}\vspace{-2mm}

We have carried out our computations on CLS lattices of size $48\times 24^3$.
The configurations have been generated with non-perturbatively $\Oa$ improved
fermions at $\beta=5.3$ and sea quark masses given by $\kappa=0.13635,0.13625$.
This roughly corresponds to $a \approx 0.08~\fm$ and $L\approx 2~\fm$,
with dynamical pion masses slightly below $300~\MeV$ and $400~\MeV$, respectively.
We will refer to these two lattices as D$_5$ and D$_6$. It has to
be noted that for the D$_6$ lattice we have two statistically independent ensembles,
that we dub D$_{6{\rm a}}$ and D$_{6{\rm b}}$. We have analysed 237 D$_6$
configurations and 137 D$_5$ configurations; in both cases successive saved
configurations are separated by 30 HMC trajectories.
Further details concerning the simulations can be obtained in~\cite{DelDebbio:2007pz}.
Our Neuberger fermion
code is the same used in previous quenched studies~\cite{rmtnum1,quenched},
and is designed specifically to perform efficiently in the
$\epsilon$-regime~\cite{Giusti:2002sm}.

\begin{figure}[!t]
\vspace*{-7mm}
\hspace*{4mm}\includegraphics[scale=0.55]{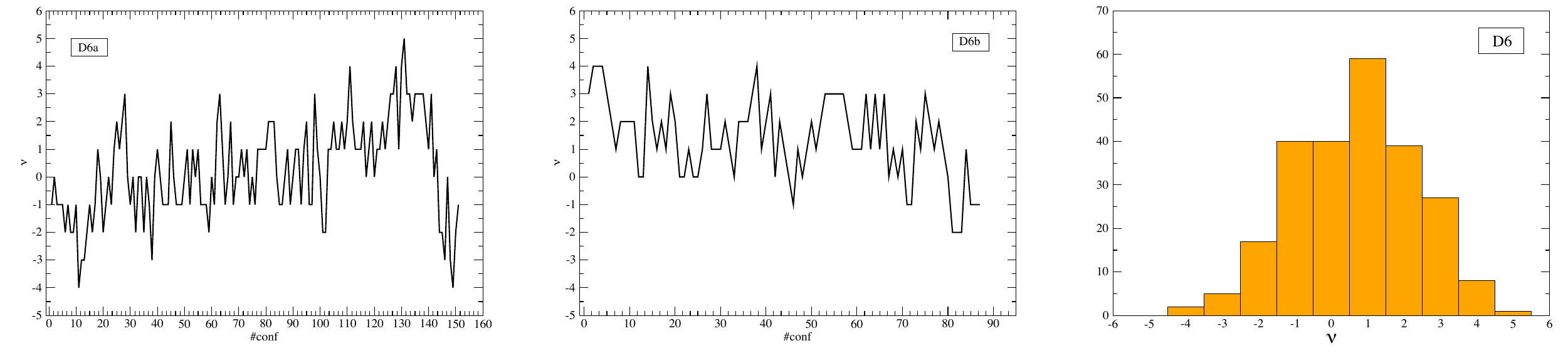}
\\\vspace{-9mm}
\caption{MC history and distribution (right panel) of the 
topological charge in D$_6$ lattices.}
\label{fig:topology}
\end{figure}

A first, immediate application of having constructed the Neuberger-Dirac operator
$D_{\rm N}$ on a given dynamical configuration is a non-ambiguous determination
of the topological
charge of the latter by computing the index of $D_{\rm N}$. In~\refig{fig:topology}
we show as an example the Monte Carlo history of the topological charge for
lattice D$_6$, which shows that topology sampling proceeds smoothly, although
the topological charge is often observed to remain constant for several tens
of trajectories. The histogram in the lower panel shows the distribution of the
measured topological charges, which exhibits the expected Gaussian-like shape and width.
This finding is consistent with the study reported in~\cite{Schaefer:2009xx}, since our computations
take place at a value of the lattice spacing sufficiently larger than the threshold
$a \sim 0.05~\fm$ below which topology is expected to exhibit freezing symptoms.

\begin{figure}[!h]
\vspace*{-2mm}
\hspace*{4mm}\includegraphics[scale=0.55]{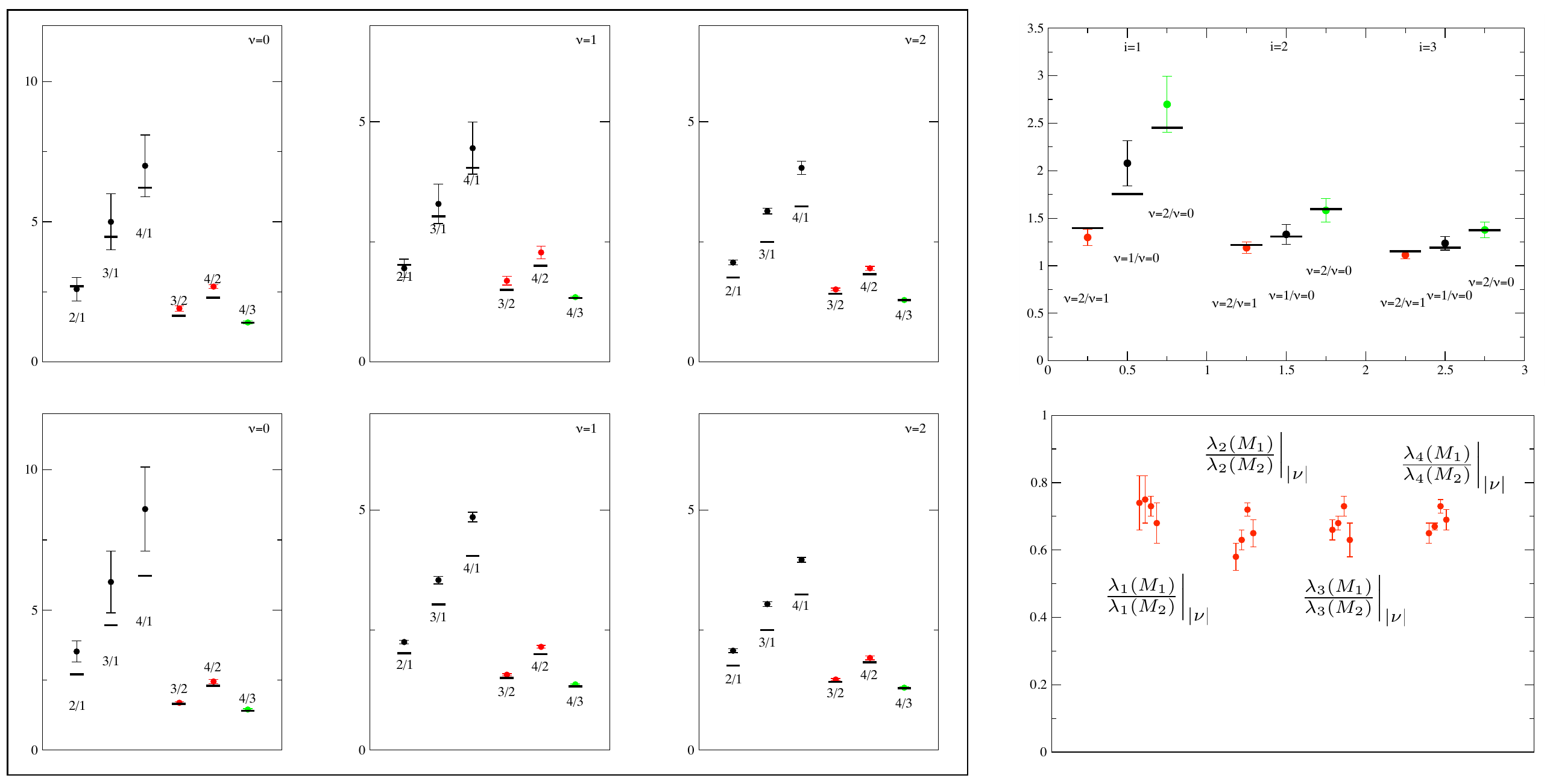}
\\\vspace{-8mm}
\caption{Left panel: Ratios of Dirac eigenvalues in different topological sectors
($k/l$ is shorthand for $\lambda_k/\lambda_l$) for lattices D$_5$ (top) and D$_6$ 
(bottom). Upper right: ratios of eigenvalues for different topologies in lattice
D$_6$.
Bottom right: ratios of D$_5$ and D$_6$ eigenvalues at fixed topology
($|\nu|=0,\ldots,3$ from left to right in each group). Horizontal ticks appearing
in plots are quenched RMT predictions.}
\label{fig:ev_ratios_all}
\end{figure}

The 10 lowest lying eigenvalues of the Dirac-Neuberger operator have been computed
on both lattices D$_5$ and D$_6$, using the techiques described in~\cite{Giusti:2002sm}. As explained
above, we expect them to be described by quenched RMT probability distributions,
with the appropriate value of the effective chiral condensate in
Eq.~(\ref{eq:condensate}). As a test, the computed ratios $\lambda_k/\lambda_l$
involving the four lowest-lying eigenvalues are compared in \refig{fig:ev_ratios_all} to quenched RMT for $|\nu|=0,1,2$. While the RMT prediction seems to work well for
ratios not involving $\lambda_1$, the ratios $\lambda_k/\lambda_1$ exhibit deviations
which are particularly noticeable in lattice D$_6$. On the other hand, ratios
between eigenvalues in different topological sectors follow well RMT predictions
also in the case of $\lambda_1$, as shown in \refig{fig:ev_ratios_all}, albeit with larger
errors. The origin
of the observed deviations, and its possible relation to chiral corrections,
will be the subject of further investigation.

In the spirit of the mixed regime ChiPT analysis, our data also allow to study
the mass dependence of the effective condensate, cf. Eq.~(\ref{eq:ev_condensate}).
\refig{fig:ev_ratios_all} shows to what precision ratios of eigenvalues computed on
gauge configurations with different dynamical pion masses do not depend neither
on topology nor on the eigenvalue number. This is a highly nontrivial test that
the sea pion mass dependence enters in the eigenvalues as predicted by our
ChiPT description. Averaging over these ratios leads to a preliminary value
$\tilde\Sigma(M_{{\rm D}_6})/\tilde\Sigma(M_{{\rm D}_5})=0.68(4)$, where the
quoted uncertainty is statistical only. The dependence on the sea pion mass,
driven by the LEC $L_6$, shows the expected sign. A determination of the
LEC itself will be the subject of a future detailed study.

\vspace{-2mm}\section{Outlook}\vspace{-2mm}

We have implemented a mixed action approach to lattice QCD in which sea quarks
are non-perturbatively $\Oa$ improved Wilson fermions, while valence quarks are
Neuberger fermions. Numerical techniques to deal with the latter that were developed
in previous quenched studies have proven similarly efficient in this context.
As a first application we have studied the Dirac spectrum in the background of
dynamical configurations at $a\approx 0.08~\fm$, and compared our findings to
expectations from mixed-regime ChiPT. The latter describe well the data, and allow
us e.g. to study the mass dependence of the chiral condensate.

In upcoming publications we will discuss results for standard two- and
three-point functions, both in the $\epsilon$- and the $p$-regime, with a view
to phenomenological applications. One specific topic that remains to be addressed
is the role of cutoff effects in the mixed action setup, and whether $\Oasq$
unitarity violating artifacts lead to sizeable scaling
violations, as discussed in~\cite{mixedactions}.

\vskip 2mm

F.B. and C.P. acknowledge financial support from the FPU grant AP2005-5201 and
the Ram\'on y Cajal Programme, respectively.
This work was partially supported by the Spanish Ministry for Education and Science projects FPA2006-05807, FPA2006-60323, FPA2008-01732, HA2008-0057 and CSD2007-00042; the Generalitat Valenciana (PROMETEO/2009/116); the Comunidad Aut\'onoma de Madrid (HEPHACOS P-ESP-00346);
and the European project FLAVIAnet (MRTN-CT-2006-035482). 
Our simulations were performed on the IBM MareNostrum at the Barcelona Supercomputing Center, as well as PC clusters and the Tirant installation at IFIC.
We thankfully acknowledge the computer resources and technical support provided by
these institutions.


\begin{thebibliography}{99}

\vspace{-1mm}\bibitem{dynrev}
See e.g. C.~Jung, PoS LAT2009 (2009) 002.

\vspace{-1mm}\bibitem{epsilon}
  J.~Gasser and H.~Leutwyler,
  Nucl.\ Phys.\  B 307, 763 (1988),
  Phys.\ Lett.\  B 188, 477 (1987);
H.~Neuberger,
  Phys.\ Rev.\ Lett.\  60 (1988) 889,
  Nucl.\ Phys.\  B 300 (1988) 180.
  
\vspace{-1mm}\bibitem{rmt}
E.V.~Shuryak, J.J.M.~Verbaarschot, Nucl. Phys. A 560 (1993) 306;
J.J.M.~Verbaarschot, I.~Zahed, Phys. Rev. Lett. 70 (1993) 3852;
J.J.M.~Verbaarschot, Phys. Rev. Lett. 72 (1994) 2531.
  
\vspace{-1mm}\bibitem{rmtnum1}
  L.~Giusti, M.~L\"uscher, P.~Weisz and H.~Wittig,
  JHEP 0311 (2003) 023.

\vspace{-1mm}\bibitem{rmtnum2}
 H.~Fukaya {\it et al.},
  Phys.\ Rev.\  D76 (2007) 054503.

\vspace{-1mm}\bibitem{dynov}
See e.g. H.~Fukaya, PoS LAT2009 (2009) 002, for an up-to-date review.
  
\vspace{-1mm}\bibitem{cls}
\verb+http://twiki.cern.ch/twiki/bin/view/CLS/WebIntro+ 

\vspace{-1mm}\bibitem{mixedreg}
  F.~Bernardoni and P.~Hern\'andez,
  JHEP 0710 (2007) 033;
  F.~Bernardoni, P.H.~Damgaard, H.~Fukaya and P.~Hern\'andez,
  JHEP 0810 (2008) 008.

\vspace{-1mm}\bibitem{zm}
  L.~Giusti, P.~Hern\'andez, M.~Laine, P.~Weisz and H.~Wittig,
  JHEP 0401 (2004) 003;
  P.~Hern\'andez {\it et al.},
  JHEP 0805 (2008) 043.

\vspace{-1mm}\bibitem{DelDebbio:2007pz}
  L.~Del Debbio, L.~Giusti, M.~L\"uscher, R.~Petronzio and N.~Tantalo,
  JHEP {\bf 0702} (2007) 082.
  
\vspace{-1mm}\bibitem{quenched}
  L.~Giusti, P.~Hern\'andez, M.~Laine, P.~Weisz and H.~Wittig,
  JHEP 0404 (2004) 013;
  L.~Giusti {\it el al.},
  Phys.\ Rev.\ Lett.\  98 (2007) 082003;
  L.~Giusti and S.~Necco,
  JHEP 0704, 090 (2007);
  L.~Giusti {\it et al.},
  JHEP 0805 (2008) 024.

\vspace{-1mm}\bibitem{Giusti:2002sm}
  L.~Giusti, C.~Hoelbling, M.~L\"uscher and H.~Wittig,
  Comput.\ Phys.\ Commun.\ 153, 31 (2003).

\vspace{-1mm}\bibitem{Schaefer:2009xx}
  S.~Schaefer, R.~Sommer and F.~Virotta, PoS LAT2009 (2009) 032.

\vspace{-1mm}\bibitem{mixedactions}
  M.~Golterman, T.~Izubuchi and Y.~Shamir,
  Phys.\ Rev.\  D71 (2005) 114508;
  S.~D\"urr {\it et al.},
  PoS LAT2007 (2007) 115.

\end{thebibliography}
\end{document}